\documentclass[review]{elsarticle}
\usepackage{epstopdf}
\usepackage{exscale}

\usepackage{relsize}
\biboptions{numbers,sort&compress}
\usepackage{booktabs} 
\usepackage{array} 
\usepackage{paralist} 
\usepackage{verbatim} 
\usepackage{subfig} 
\usepackage{bm}
\usepackage{graphicx} 
\usepackage{bm}
\usepackage{amsmath,breqn}
\usepackage{geometry} 
\usepackage{amssymb}
\usepackage{float}
\usepackage{graphicx}
\usepackage{amsfonts,amssymb}
\usepackage{amsmath}
\usepackage{booktabs} 
\usepackage{array} 
\usepackage{paralist} 
\usepackage{verbatim} 
\usepackage{subfig} 
\usepackage{geometry} 
\usepackage{color, soul}

\usepackage[pdftex,colorlinks,linkcolor=red,anchorcolor=blue,citecolor=green]{hyperref}

\usepackage{graphicx} 

\usepackage{amsmath}
\usepackage{booktabs} 
\usepackage{array} 
\usepackage{paralist} 
\usepackage{verbatim} 
\usepackage{subfig} 
\usepackage{amsthm}

\allowdisplaybreaks[4]
\usepackage{lineno,hyperref}
\modulolinenumbers[5]

\usepackage{amsfonts,amsmath,amsthm,amssymb,mathrsfs}
\usepackage{color}
\usepackage{amsmath,amssymb,amsfonts,bm}
\usepackage{graphicx}
\usepackage{newunicodechar}
\usepackage{xcolor}
\numberwithin{equation}{section}

\theoremstyle{plain}

\allowdisplaybreaks[4]

\usepackage{lineno,hyperref}
\modulolinenumbers[5]

\journal{Journal of \LaTeX\ Templates}

\journal{Journal of \LaTeX\ Templates}

\setulcolor{red}
\begin{document}
	
	\begin{frontmatter}
		
		\title{Defocusing Hirota equation with fully asymmetric non-zero boundary conditions: the inverse scattering transform}


		\author[mymainaddress]{Rusuo Ye}
		
			\author[mymainaddress]{Peng–Fei Han}

		\author[mymainaddress]{Yi Zhang\corref{mycorrespondingauthor}}
		
		\cortext[mycorrespondingauthor]{Corresponding author}

		\ead{zy2836@163.com}

		\address[mymainaddress]{ Department of Mathematics, Zhejiang Normal University, Jinhua 321004, PR China}

		\begin{abstract}	
The paper aims to apply the inverse scattering transform to the defocusing Hirota equation with fully asymmetric non-zero boundary conditions (NZBCs), addressing scenarios in which the solution's limiting values at spatial infinities exhibit distinct non-zero moduli. In comparison to the symmetric case,  we explore  the characteristic branched nature of the relevant scattering problem explicitly, instead of introducing Riemann surfaces. For the direct problem, we formulate the Jost solutions and scattering data on a single sheet of the scattering variables. We then derive their analyticity behavior, symmetry properties, and the distribution of discrete spectrum. Additionally, we study the behavior of the eigenfunctions and scattering data at the branch points.	Finally, the solutions to the  defocusing Hirota equation with asymmetric NZBCs  are presented through the related Riemann-Hilbert problem  on an open contour.  Our results
	can be applicable to the study of asymmetric conditions in nonlinear optics. 
		\end{abstract}
		
		\begin{keyword}
		Fully asymmetric non-zero boundary conditions,	Inverse scattering transform,   Riemann-Hilbert problem,  Hirota equation 
		\end{keyword}
		
	\end{frontmatter}

	\section{Introduction}
	The inverse scattering transform (IST) is an effective approach for studying integrable systems and deriving their soliton solutions. It has been extensively applied to investigate various integrable nonlinear wave equations, including the nonlinear Schr$\rm\ddot{o}$dinger (NLS) equation \cite{DC-1,c-2,c-3,DC-2}, Sasa-Satsuma equation \cite{ss-1,ss-2,ss-3}, derivative NLS equation \cite{d-1,d-2,d-3,d-4,d-5},  modified Korteweg-de Vries (mKdV) equation \cite{kdv-1,kdv-2,kdv-3,kdv-4,DZ-5,mmm} and  etc. The NLS equation, given by 
	\begin{equation}\label{ss-1}
		ip_t+p_{xx}+2\sigma |p|^2p=0,
	\end{equation}
is a commonly used model for describing weakly nonlinear dispersive waves.   Here, the values of  $\sigma=1$ and $\sigma=-1$   represent the focusing and defocusing regimes, respectively. For the focusing NLS equation,   Zakharov and Shabat firstly developed the   IST   with zero boundary conditions (ZBCs)  \cite{DC-1}, and later, Biondini and Kova$\rm\check{c}$i$\rm\check{c}$ solved the initial value problem  with non-zero boundary conditions (NZBCs)  via IST \cite{c-2}. For the defocusing NLS equation,
 the application of the IST     with   NZBCs was firstly presented by Zakharov and Shabat   \cite{DC-2} and a rigorous theory of the IST   with NZBCs was subsequently formulated by Demontis et al. \cite{c-3}. Since then,  there has been significant attention paid to the IST of numerous integrable  equations with both ZBCs and NZBCs, utilizing solutions derived from the corresponding Riemann-Hilbert problem (RHP) \cite{DZ-3,DZ-4,q-1,m-1,m-2,g-1,g-2,g-3,g-4,g-5,g-6,g-7}.
However, while there is a significant body of literature on integrable equations with  NZBCs, the results are  confined to situations where the boundary conditions are entirely symmetric.  In some physical applications, it is important to study the situations where the boundary condition is fully asymmetric. Asymmetric conditions in nonlinear optics describe a scenario where a continuous wave laser smoothly transitions between different power levels. Therefore,   it is crucial to study the integrable equations with asymmetric NZBCs. In 1982, Boiti and Pempinelly firstly investigated the defocusing NLS equation with asymmetric NZBCs   \cite{q-2}. They formulated a four-sheeted  Riemann surface, however, they did not establish the RHP, nor did they characterize the spectral data or solutions.  In 2014, Demontis et al. developed the IST to solve the initial-value problem for the focusing NLS equation with fully asymmetric NZBCs \cite{DZ-2}. Recently, Biondini et al. studied the defocusing NLS equation with fully asymmetric NZBCs \cite{DZ-1}.  The theory in \cite{DZ-1,DZ-2} is formulated without relying on Riemann surfaces, instead, it explicitly addresses the branched nature of the eigenvalues associated with the scattering problem.  To the best of our knowledge, no studies  have been conducted on the IST for the defocusing Hirota equation with fully asymmetric NZBCs.

This work is concerned the  defocusing
Hirota equation with fully asymmetric NZBCs:
	\begin{equation}\label{e1}
	\left\{
	\begin{aligned}
		&ip_t+\alpha(p_{xx}-2  |p|^2p)+i\beta(p_{xxx}-6  |p|^2p_x)=0,\quad \alpha,\beta\in\mathbb{R},\\
		&\lim_{x\to\pm\infty}{p}(x,t)={p}_{\pm}(t), \quad   |{p}_+(t)|\neq|{p}_-(t)|,\quad \hbox{arg}~p_+(t)\neq \hbox{arg}~p_-(t),
	\end{aligned}
	\right.
\end{equation}
where $p=p(x,t)$ represents the complex wave envelope. The Hirota equation is a completely integrable equation, serving as a high-order extension of the NLS  equation. It has studied  extensively  by various methods \cite{j-1,j-2,hir-1,hir-2,hir-3,hir-4,hir-5,q-1,DZ-3,r-1,r-2,r-3,r-4}. 
Among them, the utilization of IST for the Hirota equation  has attracted considerable attention. In \cite{q-1,DZ-3},  the soliton solutions of the Hirota equation were investigated under ZBCs and symmetric NZBCs. The asymptotic behavior of degenerate solitons and high-order solitons for the Hirota equation was explored in  \cite{r-3,r-4}.  Additionally, in \cite{r-2}, the Fokas method was employed to address initial-boundary-value problems for the Hirota equation on the half-line. 

In the limits $\alpha\to 0$ and $\beta\to 0$, \eqref{e1} becomes the NLS equation and mKdV equation with fully asymmetric NZBCs, respectively.  Remarkable progress has been made in IST for the mKdV equation. The solutions  with up to triple poles  of the focusing mKdV equation were studied \cite{kdv-2, kdv-3}. Later, 
Demontis derived the soliton solutions and breathers for the mKdV equation with ZBCs  \cite{kdv-4}. After that, 
the soliton solutions of  mKdV equations with  symmetric NZBCs  were also  investigated   \cite{kdv-1,kdv-2,kdv-3,kdv-4,DZ-5}.
Recently, Baldwin studied the long-time asymptotic behavior of  solution   for the focusing mKdV equation with step-like NZBCs, i.e. $p_-\neq p_+=0$ \cite{mmm}.  

Note that when the spatial derivative of $p(x,t)$ approaches zero as $x\to\pm\infty$, \eqref{e1} yields $|p_{\pm}(t)|=|p_\pm(0)|$. In this work, we choose the following boundary conditions:   
\begin{equation}\label{e2}
	p_\pm(t)=\mu_{\pm}\hbox{e}^{i\gamma_{\pm}-2i  \alpha \mu_\pm^2t},
\end{equation}
with $0\leq\gamma_\pm< 2\pi$ and $\mu_\pm>0$. Due to the  symmetry $x\mapsto -x$ and $\beta\mapsto -\beta$ for the Hirota equation, we consider   $\mu_->\mu_+>0$  without loss of generality. 

The paper is arranged as follows.
In Section 2, we introduce the direct problem, exploring the analyticity behavior, symmetry properties, and the distribution of discrete spectrum.   Section 3 is devoted to the time evolution. We determine the evolution for scattering data, reflection coefficients and norming constants.    In Section 4, we present the inverse scattering problem as a matrix RHP and obtain  the solutions for the defocusing Hirota equation with asymmetry NZBCs.
 
\section{Direct problem}
 Equation \eqref{e1} admits the following Lax pair:
\begin{equation}\label{e3}
	\Psi_x=U(x,t,z)\Psi, \quad
	 \Psi_t=V(x,t,z)\Psi,
\end{equation}
(the first of which is usually called the ``scattering problem"), where $\Psi=\Psi(x,t,z)$ and
\begin{equation}
\begin{split}
	&U=iz\sigma_3+P,\\
	&V=\alpha V_{nls}+\beta V_{cmkdv},
\end{split}
\end{equation}
with
\begin{equation}
\begin{split}
	&V_{nls}=-2zU+i\sigma_3(P_x-P^2),\\
	&V_{cmkdv}=-2zV_{nls}+[P_x,P]+2P^3-P_{xx},
\end{split}
\end{equation}
\begin{equation}
\sigma_3=\hbox{diag}(1,-1),\quad P=\left(
	\begin{array}{cc}
		0 & p(x,t) \\
		  p^*(x,t)&0\\
	\end{array}
	\right),
\end{equation}
 and the asterisk is the  complex conjugation.

The asymptotic scattering problem as $x\to\pm\infty$ of the first of \eqref{e3} is
\begin{equation}
	\Psi_x=U_\pm(z,t)\Psi,
\end{equation}
where
\begin{equation}
 U_\pm=iz\sigma_3+P_\pm,\quad P_\pm=\left(
	\begin{array}{cc}
		0 & p_\pm(t) \\
	  p_\pm^*(t)&0\\
	\end{array}
	\right).
\end{equation}
The eigenvalues of $U_\pm$ are $\pm i\lambda_\pm(z)$, where
\begin{equation}\label{e4}
	\lambda_\pm^2=z^2-  \mu_\pm^2.
\end{equation}
As in the symmetric case \cite{DZ-3}, these eigenvalues exhibit branching. In contrast to \cite{DZ-3}, the authors introduced the two-sheeted Riemann surface, here we define $\lambda_\pm$ as single-valued functions over   a single sheet of the scattering variables $\lambda_\pm=\sqrt{z^2-\mu_\pm^2}$ as in \cite{DZ-1}.

\subsection{ Jost eigenfunctions  and scattering matrix}
It will be convenient to define some notations:
\begin{equation}
	\begin{split}
	&\Xi_\pm=(-\infty,-\mu_\pm]\cup [\mu_\pm,\infty),\\
	&\Xi_{\circ}=[-\mu_-,-\mu_+]\cup [\mu_+,\mu_-],\\
	&\mathring{\Xi}_\pm=(-\infty,-\mu_\pm)\cup (\mu_\pm,\infty),	\\
	&\mathring{\Xi}_{\circ}=(-\mu_-,-\mu_+)\cup (\mu_+,\mu_-).
\end{split}
\end{equation}

	 	As $x\to\pm\infty$, the branch points are the values of $z$ for which $\lambda_\pm=0$, i.e. $z=\pm \mu_{\pm}$.  We take the branch cuts on $\Xi_\pm$ (see Fig. 1). We define $\lambda_\pm$ as   analytic functions for  all $z\in\mathbb{C}\backslash\Xi_\pm$, and these functions remain continuous as  $z$ approaches $\Xi_\pm$ from above. We see that Im$\lambda_\pm\geq 0$ and Im$(\lambda_\pm\pm z)\geq 0$ for all $z\in\mathbb{C}$.  Clearly, $\Xi_-\subset\Xi_+$ and $\lambda_\pm\in\mathbb{R}$, $\forall$  $z\in\Xi_-$. Thus,   continuous spectrum of the scattering problem  consists of $z\in\Sigma_-$.
 	
\begin{figure}[htbp]
	\centering
	\includegraphics[width=0.8\textwidth]{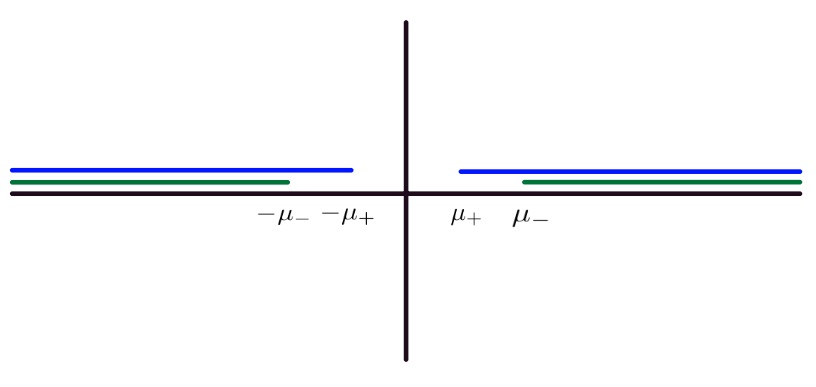}  
	\caption{The branch cuts $\Xi_-$ and $\Xi_+$ of the complex $z$-plane.}
	\end{figure}

 Similarly to \cite{DZ-3},   the eigenvector matrices of $U_\pm$ can be expressed as follows:
 \begin{equation}\label{xu2}
 	X_\pm(z,t)=I+\frac{i}{z+\lambda_\pm}\sigma_3P_\pm.
 \end{equation}
For  $z\in\mathring{\Xi}_{\pm}$, we  introduce the Jost solutions $\Psi_\pm(x,t,z)$ by
\begin{equation}\label{xu1}
	\Psi_\pm=X_\pm\hbox{e}^{i\lambda_\pm x\sigma_3}(I+o(1)),\quad x\to\pm\infty.
\end{equation}
Let
\begin{equation}	X_\pm^{-1}=\frac{1}{d_\pm(z)}[I-i\sigma_3P_\pm/(z+\lambda_\pm)],
	\end{equation}
	where
	\begin{equation}
	d_\pm(z):=\hbox{det}X_\pm=\frac{2\lambda_\pm}{z+\lambda_\pm}.
\end{equation}

We introduce the the modified eigenfunctions  
\begin{equation}\label{e7}
	v_\pm(x,t,z)=\Psi_\pm\hbox{e}^{-i\lambda_\pm x\sigma_3}.
\end{equation}
It is evident that
\begin{equation}\label{mao1}
	\lim_{x\to\pm\infty}v_\pm=X_\pm.
\end{equation}
One can formally integrate the ODE for $v_\pm$ to obtain
\begin{subequations}\label{e0}
\begin{align}
		v_-(x,t,z)=&X_-+\int_{-\infty}^xX_-\hbox{e}^{i\lambda_-(x-\xi)\hat{\sigma}_3}[X_-^{-1}\Delta P_-(\xi,t)v_-(\xi,t,z)]d\xi,\\
		v_+(x,t,z)=&X_+-\int_x^{+\infty}X_+\hbox{e}^{i\lambda_+(x-\xi)\hat{\sigma}_3}[X_+^{-1}\Delta P_+(\xi,t)v_+(\xi,t,z)]d\xi,
	\end{align}
\end{subequations}
where $\hbox{e}^{\alpha \hat{\sigma}_3}A:=\hbox{e}^{\alpha\sigma_3}A\hbox{e}^{-\alpha \sigma_3}$ and $\Delta P_\pm(x,t):=P(x,t)-P_\pm(t)$.

Let $v_\pm=(v_{\pm,1},v_{\pm,2})$. 
Using the standard Neumann iteration for   \eqref{e0}, we can prove that if $p(x,t)-p_\pm(t)\in L^1(\mathbb{R}^\pm)$, then $v_{-,2}$ is analytic in $\mathbb{C}\backslash\Xi_-$,  whereas $v_{+,1}$ is analytic in $\mathbb{C}\backslash\Xi_+$. In addition, for $t\geq 0$, when $(1+|x|)(p(x,t)-p_\pm(t))\in L^1(\mathbb{R}^\pm)$,  \eqref{e0} are well-defined when $z\to \pm \mu_\pm$. We can see that $v_-$ and $v_+$ admit the form as $z\to \pm \mu_-$ and $z\to\pm \mu_+$, respectively,
\begin{subequations}
	\begin{align}
		&v_{-}(x,t,\pm \mu_-)=I\pm i\sigma_3P_-/\mu_-+\int_{-\infty}^x[(x-\xi)U_-(\pm \mu_-,t)+I]\Delta P_-(\xi,t)v_-(\xi,t,\pm \mu_-)d\xi,\\
		&v_{+}(x,t,\pm \mu_+)=I\pm i\sigma_3P_+/\mu_+-\int_x^{+\infty}[(x-\xi)U_+(\pm \mu_+,t)+I]\Delta P_+(\xi,t)v_+(\xi,t,\pm \mu_+)d\xi.
	\end{align}
\end{subequations}

Using tr$U=0$ and Abel's formula, we find that  $\hbox{det}\Psi_\pm$ is independent of $x$. Evaluate $\hbox{det}\Psi_\pm$ as $x\to\pm\infty$ to obtain
\begin{equation}
	\hbox{det}\Psi_\pm=\hbox{det}v_\pm=d_\pm.
\end{equation}
Since both $\Psi_\pm$ solve the scattering problem for $z\in\mathring{\Xi}_-$,  one has
 \begin{equation}\label{fangxia1}
 \Psi_-=\Psi_+S(z,t),\quad z\in {\Sigma}_-,	
 \end{equation}
with 
\begin{equation}\label{e00}
	\hbox{det}S=d_-/d_+.
\end{equation}
It is mentioned that $S$ is independent of $x$.
From \eqref{e00}, we have $\hbox{det}S\neq 1$,   which is a significant distinction from the case of symmetric NZBCs  \cite{DZ-3}.  Let $S(z,t)=(s_{ij}(z,t))_{1\leq i,j\leq 2}$. Using \eqref{fangxia1},   $s_{ij}$ $(i,j=1,2)$ can be expressed as follows:
	\begin{subequations}\label{fangxia2}
\begin{align}
		&s_{11}=\hbox{det}(\Psi_{-,1},\Psi_{+,2})/d_+,\quad s_{12}=\hbox{det}(\Psi_{-,2},\Psi_{+,2})/d_+,\\
				&s_{21}=\hbox{det}(\Psi_{+,1},\Psi_{-,1})/d_+,\quad s_{22}=\hbox{det}(\Psi_{+,1},\Psi_{-,2})/d_+.
\end{align}
 	\end{subequations}
From \eqref{fangxia2}, it is shown that $s_{22}$ is analytic for $z\in\mathbb{C}\backslash\Xi_+$. Because $\Psi_{-,2}$ can be extended analytically to $z\in\mathbb{C}\backslash\Xi_-$, $\Psi_{+,1}$ and $\Psi_{+,2}$ are defined on $\Xi_+$, we may extend the definitions of $s_{12}$ and $s_{22}$ pointwise to  $\mathring{\Xi}_+$.  Since $d_+$ has a double zero at $z=\pm \mu_+$, $s_{12}$ and $s_{22}$ are singular at $z=\pm \mu_+$ .

It will be useful  to  introduce the reflection coefficients
 \begin{equation}\label{yangwan1}
 r(z,t)=s_{21}/s_{11},\quad \hat{r}(z,t)=s_{12}/s_{22},\quad z\in \Xi_-,
 \end{equation}
 which will be needed in the following discussions.
 \subsection{Symmetries}
 Due to the involutions $z\mapsto z^*$ and $\lambda_\pm(z)\mapsto -\lambda_\pm(z)$, we have the two   kinds of symmetries.

(i) The first symmetry follows from $z\mapsto z^*$. It can be directly verified  that  $\sigma_1\Psi^*(x,t,z^*)\sigma_1$  also satisfies the scattering problem and  demonstrates identical asymptotic behavior to $\Psi(x,t,z)$ as
 $x\to\pm\infty$, where
 \begin{equation*}
 	\sigma_1=\left(
 	\begin{array}{cc}
 		0 & 1 \\
 		1&0\\
 	\end{array}
 	\right).
 \end{equation*} 
Hence, we  have
\begin{equation}\label{e9}
	\Psi_\pm=\sigma_1\Psi_\pm^*\sigma_1,\quad z\in\Xi_\pm.
\end{equation}

 Combining the \eqref{fangxia1} and \eqref{e9}, we have the following symmetry relation
 \begin{equation}
 	S=\sigma_1S^*\sigma_1,\quad z\in\Xi_-,
 \end{equation}
which yields
\begin{equation}\label{fangxia5}
	s_{11}=s_{22}^*,\quad s_{21}=s_{12}^*,\quad z\in\Xi_-,
\end{equation}
and 
\begin{equation}\label{fs-1}
r=\hat{r}^*,\quad z\in\Xi_-.
\end{equation}
 It follows from \eqref{e00} and \eqref{fangxia5} that
 \begin{equation}
 	\hbox{det}S=|s_{22}|^2-|s_{12}|^2=d_-/d_+,\quad z\in\Xi_-.
 \end{equation}
From $d_-/d_+>0$ for $z\in\mathring{\Xi}_-$, we see  that $s_{22}$ has no zeros on $z\in\mathring{\Xi}_-$.

Next, we consider $z\mapsto z^*$ for $z\notin \Sigma_+$. One can verify that   if $\Psi(x,t,z)$ satisfies the scattering problem, then $\sigma_1\Psi^*(x,t,z^*)$ also satisfies it. Taking the limit $x\to -\infty$, we have
 \begin{equation}\label{deng-1}
 	\sigma_1\Psi_{-,2}^*(x,t,z^*)=\frac{-ip^*_-}{z-\lambda_-}\Psi_{-,2}(x,t,z), \quad z\notin\Xi_+.
 \end{equation}
Similarly,   take the limit as  $x\to\infty$ to obtain  
\begin{equation}\label{deng-2}
 \sigma_1\Psi_{+,1}^*(x,t,z^*)=\frac{ip_+}{z-\lambda_+}\Psi_{+,1}(x,t,z),  \quad z\notin\Xi_+. 
\end{equation}
Using the formula \eqref{fangxia2}, we find that
\begin{equation}\label{deng-3}
	s_{22}^*(z^*,t)=\frac{p_+}{p_-}\frac{z+\lambda_-}{z+\lambda_+}s_{22}(z,t),\quad z\notin\Xi_+.
\end{equation}
 
(ii) The second symmetry arises from an alternative selection of $\lambda_\pm(z)$, i.e. $\lambda_\pm(z)\mapsto -\lambda_\pm(z)$. To prevent any confusion  caused by notation, we denote
\begin{equation}
	\hat{\lambda}_\pm(z):=-\lambda_\pm(z).
\end{equation}
 It is worth noting that this choice does not affect the formal independence of the integral equations for the eigenfunctions.
If $\Psi_\pm(x,t,z,\lambda_\pm(z))$ represents the solution  to the scattering problem, it follows that $\hat{\Psi}_\pm(x,t,z):=\Psi_\pm(x,t,z,\hat{\lambda}_\pm(z))$  is also a valid solution.

From \eqref{xu2} and \eqref{xu1}, we define the Jost solutions $\hat{\Psi}_\pm(x,t,z)$ admit the following asymptotic behavior
\begin{equation}
	\hat{\Psi}_\pm=(I+\frac{i}{z-\lambda_\pm}\sigma_3P_\pm)\hbox{e}^{-i\lambda_\pm x\sigma_3}(I+o(1)),\quad x\to\pm\infty.
\end{equation}
 Since $\Psi_\pm$ and $\hat{\Psi}_\pm$ are matrix solutions for the first part of the \eqref{e3} for all $ z\in\mathring{\Xi}_\pm$, we express
	\begin{subequations}\label{shenxianj}
	\begin{align}
		&\hat{\Psi}_\pm=\Psi_\pm\frac{i}{z-\lambda_\pm}\sigma_3P_\pm,\quad z\in\mathring{\Xi}_\pm,\\
		&{\Psi}_\pm=\hat{\Psi}_\pm\frac{i}{z+\lambda_\pm}\sigma_3P_\pm,\quad z\in\mathring{\Xi}_\pm.
	\end{align}
\end{subequations}
  
Define $\hat{S}(z,t)$ as the scattering matrix for $\hat{\Psi}_\pm$. Following direct calculations, we have
 \begin{equation}
 	S=\frac{z-\lambda_-}{z-\lambda_+}\sigma_3P_+\hat{S}P_-^{-1}\sigma_3,\quad z\in\Xi_-.
 \end{equation}
 Thus, elements in $S$ are simply related to  elements in $\hat{S}$ as
 \begin{equation}
 	s_{22}=\frac{p_+^*}{p_-^*}\frac{z-\lambda_-}{z-\lambda_+}\hat{s}_{11},\quad s_{21}=-\frac{p_+^*}{p_-}\frac{z-\lambda_-}{z-\lambda_+}\hat{s}_{12},\quad z\in\Xi_-.
  \end{equation}

  According to the definition of $\lambda_\pm$,  it is clear that $\lambda_\pm$ are defined to be continuous as $z\to\Xi_\pm$ from above, i.e.
\begin{equation}
	\lambda_\pm^+(z):=\lim_{\epsilon\downarrow 0}\lambda_\pm(z+i\epsilon)=\lambda_\pm.
\end{equation} 
And, $\lambda_\pm^-$ as $z\to\Xi_\pm$ from below, are given by 
\begin{subequations}
	\begin{align}
	&\lambda_\pm^-(z):=\lim_{\epsilon\uparrow 0}\lambda_\pm(z+i\epsilon)=\hat{\lambda}_\pm,\quad z\in\Xi_-,\\
	&\lambda_-^-(z):=\lim_{\epsilon\uparrow 0}\lambda_-(z+i\epsilon)=\lambda_-,\quad z\in\Xi_\circ,\\
	&\lambda_+^-(z):=\lim_{\epsilon\uparrow 0}\lambda_+(z+i\epsilon)=\hat{\lambda}_+,\quad z\in\Xi_\circ.
\end{align}
\end{subequations}
Using the definition \eqref{e7} and analytical properties of $v_{-,2}$ and $v_{+,1}$, we see that $\Psi_{-,2}$ is analytic for $z\in\mathbb{C}\backslash\Xi_-$ and exhibits continuity towards  $\Xi_-$ from above, and $\Psi_{+,1}$ is analytic for $z\in\mathbb{C}\backslash\Xi_+$ and exhibits continuity towards $\Xi_+$ from above. On the other hand, $\Psi_{-,2}$ and $\Psi_{+,1}$  as $z\to\Xi_\pm$ from below, are given by 
\begin{subequations}\label{d1}
	\begin{align}
		\Psi_{-,2}^-&:=\lim_{\epsilon\uparrow 0}\Psi_{-,2}(x,t,z+i\epsilon)\\
		&=\left\{
		\begin{aligned}
			&\hat{\Psi}_{-,2},\quad z\in\Xi_-, \\
			&{\Psi}_{-,2},\quad z\in\Xi_\circ,
		\end{aligned}
		\right.\\
			\Psi_{+,1}^-&:=\lim_{\epsilon\uparrow 0}\Psi_{+,1}(x,t,z+i\epsilon)\\
			&=
			\hat{\Psi}_{+,1},\quad z\in\Xi_+.
	\end{align}
\end{subequations}
Using the relation \eqref{shenxianj}, we have
 \begin{subequations}\label{d2}
 	\begin{align}
 	&\Psi_{-,2}^-=\frac{ip_-}{z-\lambda_-}\Psi_{-,1},\quad z\in\Xi_-,\\
 	&\Psi_{+,1}^-=\frac{-ip_+^*}{z-\lambda_+}\Psi_{+,2},\quad z\in\Xi_+.
 	\end{align}
 \end{subequations}
From above relations, we  get the limits of $s_{22}$ as $z\to \Xi_\pm$ from below:
 \begin{equation} \label{d3}
 s_{22}^-:=\lim_{\epsilon\uparrow 0}s_{22}(z+i\epsilon,t)=\left\{
 	\begin{aligned}
 		&\frac{p_-}{p_+}\frac{z-\lambda_+}{z-\lambda_-}s_{11},\quad z\in\Xi_-, \\
 		&\frac{-ip_+^*}{z+\lambda_+}s_{12},\quad z\in\mathring{\Xi}_\circ.
 	\end{aligned}
 	\right.
 \end{equation}
 
 \subsection{Behavior of the scattering data at the branch points}
Recall that $\Psi_-$ is well-defined as $z\to \pm \mu_-$ and $\Psi_+(x,t,\pm \mu_{-})$ solves the scattering problem,  we see immediately that the scattering coefficients $s_{ij}$ are well defined at $z=\pm \mu_-$.  When $z=\pm \mu_-$, $d_-(z)=0$. It follows that det$S(\pm \mu_-,t)=0$ and  the columns of $\Psi_-(x,t,\pm \mu_-)$ are linearly dependent.  By  utilizing the asymptotics of  $\Psi_{-}(x,t,\pm \mu_-)$   as well as Wronskian definitions \eqref{fangxia2}, we have
\begin{equation}\label{fangxia9}
	\begin{split}
		s_{22}(\pm \mu_-,t)=\pm i\hbox{e}^{i\gamma_--2i\alpha \mu_-^2t}s_{21}(\pm \mu_-,t),\\
			s_{12}(\pm \mu_-,t)=\pm i\hbox{e}^{i\gamma_--2i\alpha \mu_-^2t}s_{11}(\pm \mu_-,t).
	\end{split}
\end{equation}
Then in view of the expressions \eqref{fangxia5}  and \eqref{fangxia9}, we find that
\begin{equation}
		 |s_{22}(\pm \mu_-,t)|=|s_{12}(\pm \mu_-,t)|\neq 0.
\end{equation}

From   \eqref{yangwan1}, we have
 \begin{equation}\label{xy2}
 |\hat{r}(\pm \mu_-,t)|=1.
 \end{equation}
Next, we consider $z\to\pm \mu_+$.  From the definitions and properties of $\Psi_\pm$ as well as the Wronskian relations \eqref{fangxia2}, we  find that only scattering coefficients $s_{22}$ and $s_{12}$ are defined for $z\in\mathring{\Xi}_{\circ}$. Since $d_+$ has a double zero at $z=\pm \mu_+$,  $s_{22}$ and $s_{12}$ have a double pole as $z\to \pm \mu_+$.
 Specifically, we obtain the limits of $s_{22}$ and $s_{12}$ as $z\to\pm \mu_+$:
\begin{equation}
	\begin{split}
		s_{22}(z,t)\to(\frac{1}{2}+\frac{(\pm \mu_+)^{1/2}}{2\sqrt{2}(z\mp \mu_+)^{1/2}}+O(z\mp \mu_+)^{1/2})\hbox{det}(\Psi_{+,1},\Psi_{-,2})(x,t,\pm \mu_+),~z\to\pm \mu_+,\\
		s_{12}(z,t)\to(\frac{1}{2}+\frac{(\pm \mu_+)^{1/2}}{2\sqrt{2}(z\mp \mu_+)^{1/2}}+O(z\mp \mu_+)^{1/2})\hbox{det}(\Psi_{-,2},\Psi_{+,2})(x,t,\pm \mu_+),~z\to\pm \mu_+.\\
	\end{split}
\end{equation}
Notice that
\begin{equation}\label{xy3}
	\lim_{z\to\pm \mu_+}\hat{r}(z,t)=\mp i\hbox{e}^{i\gamma_+-2i\alpha \mu_+^2t}.
\end{equation}
It follows that $|\hat{r}(\pm \mu_+,t)|=1$.
 \subsection{Discrete spectrum}
 Following the similar analysis in \cite{DZ-1}, we conclude that no zeros in the inverse scattering problem for $\mathring\Xi_\circ$. Specifically, 
\begin{equation}\label{kk}
	s_{22}(z,t)s_{12}(z,t)\neq 0,\quad \forall z\in\mathring\Xi_\circ,
\end{equation}
which shows that $\hat{r}(z,t)$  has no zeros  in $\mathring\Xi_{\circ}$. 
  In the following, we make the assumption that there exists a finite number of zeros of  $s_{22}(z,t)$ lie in $(-\mu_+,\mu_+)$. This condition is satisfied if $s_{22}(\pm \mu_+,t)\neq 0$.   
  
Let $z_1,\cdots,z_W$ represent the zeros of $s_{22}(z,t)$ in $(-\mu_+,\mu_+)$. At $z=z_l$, we get
   \begin{equation}\label{xiang-1}
   	\Psi_{-,2}(x,t,z_l)=b_l(t)\Psi_{+,1}(x,t,z_l),\quad l=1,2,\cdots,W,
   \end{equation}
where $b_l$ is a scalar independent of $x$ and $z$. Now, we define the norming constants
 \begin{equation}
 	c_l(t)=b_l(t)/s_{22}'(z_l,t),\quad l=1,2,\cdots, W,
 \end{equation}
 where $'$ indicates the derivative with respect to $z$. 
    
   From \eqref{deng-1}, \eqref{deng-2} and \eqref{deng-3}, one can derive the following relations:
    \begin{equation}
    	b_l=-\frac{p_+}{p_-^*}\frac{z_l-\lambda_{-,l}}{z_l-\lambda_{+,l}}b_l^*,\quad l=1,2,\cdots, W,
    \end{equation}
and
\begin{equation}
	[s_{22}'(z_l,t)]^*=\frac{p_+}{p_-}\frac{z_l+\lambda_{-,l}}{z_l+\lambda_{+,l}}s_{22}'(z_l,t),\quad l=1,2,\cdots,W,
\end{equation}
which yields
\begin{equation}
	c_l^*=-\frac{p_+^*}{p_+}c_l,\quad l=1,2,\cdots,W.
\end{equation}
\section{Time evolution}
Recall that the Jost solutions $\Psi_\pm$ defined by \eqref{xu1} do not satisfy the second equation of the Lax pair. However, due to the compatibility condition of the Lax pair, which is represented by the Hirota equation, there must exist   solution $\Phi$ that simultaneously satisfies the scattering problem and time evolution.  Now we express $\Phi_\pm$ in terms of $\Psi_\pm$ using matrices $D_\pm(z,t)$ that are independent of $x$:
\begin{equation}\label{deng-4}
	\Phi_\pm=\Psi_\pm D_\pm,
\end{equation}
which yields
\begin{equation}
(D_\pm)_t=Y_\pm(z,t)D_\pm,
\end{equation}
where
\begin{equation}
	Y_\pm=\Psi_\pm^{-1}[V\Psi_\pm-(\Psi_\pm)t].
\end{equation}
By using \eqref{xu1}, we can evaluate $Y_\pm$ as $x\to\pm\infty$ :
\begin{equation}\label{deng-5}
	Y_\pm=\lim_{x\to\pm\infty}\Psi_\pm^{-1}[V\Psi_\pm-(\Psi_\pm)t]=ig_\pm(z)\sigma_3,\quad z\in\mathring\Xi_{\pm},
\end{equation}
where $g_\pm(z)=-2\alpha z\lambda_\pm(z)-\alpha \mu_{\pm}^2+4\beta z^2\lambda_\pm(z)+2\beta\lambda_\pm(z)\mu_\pm^2.$ It follows that
\begin{equation}\label{deng-6}
	(\Psi_\pm)_t=V\Psi_\pm-\Psi_\pm Y_\pm, \quad z\in\Xi_\pm.
\end{equation}
By  utilizing  \eqref{fangxia1} and \eqref{deng-5},   we   derive 
\begin{equation}\label{deng-7}
	S_t=Y_+S-SY_-,\quad z\in\mathring\Xi_-.
\end{equation}
By substituting \eqref{deng-5} into \eqref{deng-7}, we  have
\begin{subequations}
	\begin{align}
		&s_{12}(z,t)=s_{12}(z,0)\hbox{e}^{i(g_+(z)+g_-(z))t},\quad z\in\mathring\Xi_-,\label{ws-1}\\
			&s_{22}(z,t)=s_{22}(z,0)\hbox{e}^{-i(g_+(z)-g_-(z))t},\quad z\in\mathring\Xi_-.\label{ws-2}
	\end{align}
\end{subequations}
Note that \eqref{ws-2} can  be extended in cases where $s_{22}(z,0)$ is analytic. Additionally, by making use of  \eqref{deng-6} and   \eqref{fangxia2}, \eqref{ws-2} can   be extended to $z\in\mathring\Xi_+$. Consequently, we obtain
\begin{equation}
	\hat{r}(z,t)=\hat{r}(z,0)\hbox{e}^{2ig_+(z)t},\quad z\in\Xi_+.
\end{equation}
From \eqref{ws-1}, we can deduce that at $z=z_l$,
\begin{equation}\label{xiang-2}
	s_{22}'(z_l,t)=s_{22}'(z_l,0)\hbox{e}^{-i(g_+(z_l)-g_-(z_l))t},\quad l=1,2,\cdots, W,
\end{equation}
where $'$  represents differentiation with respect to $z$.

Equation \eqref{deng-6} implies the following expressions for the derivatives of $\Psi_{-,2}$ and $\Psi_{+,1}$ with respect to $t$:
\begin{subequations}
	\begin{align}
		&(\Psi_{-,2})_t=V\Psi_{-,2}+ig_-(z)\Psi_{-,2},\\
		&(\Psi_{+,1})_t=V\Psi_{+,1}-ig_+(z)\Psi_{+,1},\quad z\in\mathring\Xi_-.
	\end{align}
\end{subequations}
By putting these equations into \eqref{xiang-1}, we can derive the time evolution of  $b_l$ as follows:  
\begin{equation}
	b_l=b_{l0}\hbox{e}^{i(g_+(z_l)+g_-(z_l))t},\quad l=1,2,\cdots,W,
\end{equation}
where $b_{l0}=b_l(0)$.  From \eqref{xiang-2}, we  have
\begin{equation}
	c_l=c_{l0}\hbox{e}^{2ig_+(z_l)t},\quad l=1,2,\cdots,W,
\end{equation}
where $c_{l0}=c_l(0)$.  Note  that $\hbox{Im}[g_\pm(z)]\neq 0$ for all $z\notin\Xi_\pm$. With $z$ fixed, there  may be  sectors where $s_{22}(z,t)\to 0$  and others where $s_{22}(z,t)\to\infty$ as $t\to\infty$. On the other hand, with  $t$ fixed,  since
 \begin{equation}
 	\lambda_\pm(z)=z-\frac{\mu_\pm^2}{2z}+O(\frac{1}{z^3}),\quad z\to\infty,
 \end{equation}
 which yields
 \begin{equation}
 	\begin{split}
 		g_+(z)-g_-(z)&=\alpha(\mu_-^2-\mu_+^2)+2z(2\beta z-\alpha)(\lambda_+(z)-\lambda_-(z))+2\beta(\mu_+^2\lambda_+(z)-\mu_-^2\lambda_-(z))\\
 		&=O(\frac{1}{z}),\quad z\to\infty,
 	\end{split}
 \end{equation}
the behavior of
$s_{22}(z,t)$ as $z\to\infty$ remains unaffected by this time dependence.

 \section{Inverse problem}
 In the following, we will establish the associated RHP on an open contour and reconstruct the solution to the defocusing Hirota equation with fully NZBCs.

 \subsection{ Matrix Riemann-Hilbert problem}
Based on the  previous analysis,    let us introduce the meromorphic matrix: 
 \begin{equation}\label{zhang-1}
 	m(x,t,z)=(v_{+,1}, \frac{v_{-,2}}{s_{22}}),\quad z\notin\Xi_+.
 \end{equation}
Note that  the definition of the projection of $m$ onto the cut from above or below is different. In particular,
 \begin{subequations}
 	\begin{align}
 		m^+&:=\lim_{\epsilon\downarrow 0}m(x,t,z+i\epsilon)\notag\\
 		&=(v_{+,1}, \frac{v_{-,2}}{s_{22}}),\quad z\in\Xi_+,\\
 		m^-&:=\lim_{\epsilon\uparrow 0}m(x,t,z+i\epsilon)\notag\\&=\left\{
 		\begin{aligned}
 			&(v_{+,1}^-, \frac{v_{-,2}^-}{s_{22}^-}),\quad z\in\Xi_-, \\
 			&(v_{+,1}^-, \frac{v_{-,2}}{s_{22}^-}),\quad z\in{\Xi}_\circ.
 		\end{aligned}
 		\right.
 	\end{align}
 \end{subequations}
 The continuity properties of the  columns of $v_\pm$   can be easily deduced from those  $\Psi_\pm$ by using \eqref{e7}.  

Now we consider the RHP on the $\Xi_+$:
\begin{equation}\label{so-1}
	m^+=m^-J(x,t,z),\quad z\in\Xi_+,
\end{equation}
with
	\begin{equation}\label{so-2}
	J=\left\{
	\begin{aligned}
		&J_{\Xi_-}(x,t,z),\quad z\in\Xi_-,\\
		&J_{\Xi_\circ}(x,t,z),\quad z\in\Xi_\circ.
	\end{aligned}
	\right.
\end{equation}
Based on the discussions in  Section 2.3, we get the behavior of  matrix $m$ as $z\to\mu_\pm$:
	\begin{equation}\label{ff-1}
		\begin{split}
	&m=O(1),\quad z\to\pm \mu_-,\\
&m=(O(1),O(z\mp \mu_+)^{1/2}),\quad z\to \pm \mu_+.	
\end{split}
\end{equation}

Next, we will calculate the jump matrices $J_{\Xi_-}$ and $J_{\Xi_\circ}$ separately. It is worth noting that we will demonstrate that the continuity of $J$  as $z\to\pm \mu_-$ and as $z\to\pm \mu_+$.

{Jump matrix for $z\in\Xi_-$.}   We use \eqref{d2} and \eqref{d3} to express $\Psi_{-,1}$ and $\Psi_{+,2}$ in terms of $\tilde{\Psi}_{-,2}$ and $\tilde{\Psi}_{+,1}$,  resulting in the following expressions:
 \begin{align}\label{kaixue}
 	&\Psi_{+,1}=\frac{-i}{z+\lambda_+}[p_+\hat{r}^*\Psi^-_{+,1}+p_+^*\frac{\Psi^-_{-,2}}{s^-_{22}}],\\
 &\frac{\Psi_{-,2}}{s_{22}}=\frac{i}{z+\lambda_+}[p_+(1-|\hat{r}|^2)\Psi^-_{+,1}-p_+^*\hat{r}\frac{\Psi^-_{-,2}}{s^-_{22}}],
 \end{align}
which  can be expressed in   form
 \begin{equation}\label{ha}
(\Psi_{+,1}, \frac{\Psi_{-,2}}{s_{22}})=(\Psi^-_{+,1}, \frac{\Psi^-_{-,2}}{s^-_{22}})\frac{i}{z+\lambda_+}\sigma_3P_+J_{\hat{r}}(z,t),\quad z\in\Xi_-,
 \end{equation}
where
 \begin{equation}
 	J_{\hat{r}}=\left(
 	\begin{array}{cc}
 		1 & \hat{r} \\
 	-\hat{r}^* &1-|\hat{r}|^2\\
 	\end{array}
 	\right).
 \end{equation}
  Then, by the \eqref{so-1} with \eqref{so-2}, we have 
 \begin{equation}\label{p1}
	J_{\Sigma_-}=(X_+-I)\left(
	\begin{array}{cc}
		\hbox{e}^{i\lambda_-x} & 0 \\
		0 &\hbox{e}^{-i\lambda_+x}\\
	\end{array}
	\right)J_{\hat{r}}\left(
	\begin{array}{cc}
		\hbox{e}^{-i\lambda_+x} & 0 \\
		0 &\hbox{e}^{i\lambda_-x}\\
	\end{array}
	\right).
\end{equation}
 
\emph{Jump matrix for $z\in\Xi_\circ.$} From  \eqref{d3}, we obtain
 \begin{equation}
\frac{\Psi_{-,2}}{s_{22}}=\frac{\Psi_{-,2}}{s_{22}^-}\frac{-ip_+^*}{z+\lambda_+}\hat{r},\quad z\in\mathring\Xi_\circ.
 \end{equation}
 By using \eqref{d2}, we arrive at \eqref{ha}, where  $\Psi_{-,2}^-=\Psi_{-,2}$, and 
\begin{equation}
J_{\hat{r}}=\left(
\begin{array}{cc}
	1 & \hat{r} \\
	-\frac{1}{\hat{r}} &0\\
\end{array}
\right),\quad z\in\Xi_\circ.
\end{equation}
 Then, by the \eqref{so-1} with \eqref{so-2}, we have 
\begin{equation}\label{p2}
	J_{\Sigma_\circ}=(X_+-I)\left(
	\begin{array}{cc}
		\hbox{e}^{-i\lambda_-x} & 0 \\
		0 &\hbox{e}^{-i\lambda_+x}\\
	\end{array}
	\right)J_{\hat{r}}\left(
	\begin{array}{cc}
		\hbox{e}^{-i\lambda_+x} & 0 \\
		0 &\hbox{e}^{i\lambda_-x}\\
	\end{array}
	\right).
\end{equation}
To express $J_r(z,t)$ over $\Xi_+$, we can use the following formula:
\begin{equation}\label{xy4}
	J_{\hat{r}}=\left(
	\begin{array}{cc}
		1 & \hat{r} \\
		-{r} &1-\hat{r}{r}\\
	\end{array}
	\right),\quad z\in\Xi_+,
\end{equation}
where we formally define
\begin{equation}\label{xy1}
r=\frac{1}{\hat{r}},\quad z\in\Xi_\circ.
\end{equation}
  Moreover, using \eqref{xy2}, we have $\hat{r}^*(\pm \mu_-,t)=1/\hat{r}(\pm \mu_-,t)$. This definition of the extended $r$ ensures its continuity at $z=\pm \mu_-$. Furthermore, from equation \eqref{xy3}, we see that $\hat{r}$ and $r$ (and thus $J_{\hat{r}}$)  are continuous for  $z\in\Xi_+$, including at  $z=\pm \mu_\pm$.

  To summarize the above results, the  RHP is formulated as follows:
  \begin{equation}
  	m^+=m^-(X_+-I)(I-J_0),\quad z\in\Xi_+,
  \end{equation}
  where
  \begin{equation}\label{n-5}
  	J_0=\left\{
  	\begin{aligned}
  		&\left(
  		\begin{array}{cc}
  			1-\hbox{e}^{-i(\lambda_+-\lambda_-)x} & -\hat{r}\hbox{e}^{2i\lambda_-x} \\
  			\hat{r}^*\hbox{e}^{-2i\lambda_+x} &1-\hbox{e}^{-i(\lambda_+-\lambda_-)x}(1-|\hat{r}|^2)\\
  		\end{array}
  		\right),\quad z\in\Xi_-,\\
  		&\left(
  		\begin{array}{cc}
  			1-\hbox{e}^{-i(\lambda_++\lambda_-)x} & -\hat{r}  \\
  			\hbox{e}^{-2i\lambda_+x}/\hat{r}&1\\
  		\end{array}
  		\right),\quad z\in\Xi_\circ.
  	\end{aligned}
  	\right.
  \end{equation}

\subsection{Asymptotic behavior}
Now, we  explore the asymptotic behavior of the Jost solutions and scattering data as $z\to\infty$. A direct calculation shows
\begin{equation} 
	\lambda_-(z)=\left\{
	\begin{aligned}
		&z-\frac{\mu_-^2}{2z}+O(1/z^2),\quad z\to\infty\wedge\hbox{Im}z\geq 0,\\\
		&-z+\frac{\mu_-^2}{2z}+O(1/z^2),\quad z\to\infty\wedge\hbox{Im}z<0.
	\end{aligned}
	\right.
\end{equation}
Now we will demonstrate that  if  $q_x(\cdot,t)\in L^1(\mathbb{R})$, then $v_{-,2}$ and $v_{+,1}$ enjoy  the following asymptotic behavior as $z\to\infty$:
	\begin{align}
&	v_{-,12}=\frac{ip}{2z}+O(1/z^2),\\
&v_{-,22}=1+O(1/z),\quad z\to\infty,\quad\hbox{Im}z\geq 0,	
	\end{align}
and 
\begin{align}
&	v_{-,12}=\frac{2iz}{p_-^*}+O(1),\\
&v_{-,22}=\frac{p^*}{p_-^*}+O(1/z),\quad z\to\infty,\quad\hbox{Im}z<0.
\end{align}
Moreover,
	\begin{align}
	&	v_{+,11}=1+O(1/z),\\
	&v_{+,21}=-\frac{ip^*}{2z}+O(1/z^2),\quad z\to\infty,\quad\hbox{Im}z\geq 0,	
\end{align}
and 
\begin{align}
	&	v_{+,11}=\frac{p}{p_+}+O(1/z),\\
	&v_{+,21}=-\frac{2iz}{p_+}+O(1),\quad z\to\infty,\quad\hbox{Im}z< 0.
\end{align}
Combing the above expressions with \eqref{fangxia2}, we obtain 
\begin{subequations}
	\begin{align}
		&s_{22}=1+O(1/z),\quad z\to\infty\wedge\hbox{Im}z>0,\\
		&s_{22}=\frac{p_+^*}{p_-^*}+O(1/z),\quad z\to\infty\wedge\hbox{Im}z<0.
	\end{align}
\end{subequations}

\subsection{Solution of the RHP  }
 Evaluating the asymptotic behaviors of $m$ as $z\to\infty$, we have 
\begin{equation}\label{n-3}
	m=\left\{
	\begin{aligned}
		&I+O(1/z)_,\quad &z\to\infty\wedge\hbox{Im}z> 0,\\\
		&\frac{i}{z+\lambda_+}\sigma_3P_++O(1),\quad &z\to\infty\wedge\hbox{Im}z<0.
	\end{aligned}
	\right.
\end{equation}
 To get a simpler jump matrix, we introduce a   matrix $m_*(x,t,z)$ and  arrive at a new RHP:
\begin{equation}\label{n-2}
m_*^+=m_*^-(X_+-I),\quad z\in\Xi_+.
\end{equation} 
A solution to this problem can be easily found by inspection, i.e.  $m_*=X_+$. We rewrite as
\begin{equation}\label{gl-1}
	X_+=X_+^-(X_+-I).
\end{equation}

Based on our analysis,  matrix $m$ can be expressed as:
\begin{equation}\label{n-4}
	m=w(x,t,z)X_+.
\end{equation}
where $w=I+O(1/z)$ as $z\to\infty$. This implies
\begin{equation}
w^+=w^-\tilde{J}(x,t,z),\quad z\in\Xi_+,
\end{equation}
where $\tilde{J}=X_+^-JX_+^{-1}$.   From \eqref{so-1}, \eqref{n-5}  and  \eqref{gl-1}, we have  
\begin{equation}\label{gl-2}
	\tilde{J}=X_+(I-J_0)X_+^{-1},\quad z\in\Xi_+,
\end{equation}
where $J_0$ is given by \eqref{n-5}. Then from \eqref{ff-1}, we have   $w=O(1)$ as $z\to\pm \mu_-$ and $w=(O(1),O(z\mp \mu_+)^{1/2})$ as $z\to \pm \mu_+$.

 From \eqref{xiang-1}, we derive
\begin{equation}
v_{-,2}(x,t,z_l)=b_lv_{+,1}(x,t,z_l)\hbox{e}^{i(\lambda_{-,l}+\lambda_{+,l})x},
\end{equation}
for $l=1,2,\cdots,W$. Because the zeros of $s_{22}(z,t)$ are simple,
\begin{equation}
	\begin{split}
\mathop{\hbox{Res}}\limits_{z=z_l}[\frac{v_{-,2}(x,t,z)}{s_{22}(z,t)}]&=\frac{v_{-,2}(x,t,z_l)}{s'_{22}(z_l,t)}\\
&=c_l\hbox{e}^{i(\lambda_{-,l}+\lambda_{+,l})x}v_{+,1}(x,t,z_l),\quad l=1,2,\cdots,W,
\end{split}
\end{equation}
where  $\lambda_{\pm,l}=\lambda_\pm(z_l)$. Therefore
\begin{equation}
\mathop{\hbox{Res}}\limits_{z=z_l}[m(x,t,z)]=c_l\hbox{e}^{i(\lambda_{-,l}+\lambda_{+,l})x}(0,m_1(x,t,z_l)),\quad l=1,2,\cdots,W,
\end{equation}
which yields
\begin{equation}\label{dengy-1}
	\begin{split}
	\mathop{\hbox{Res}}\limits_{z=z_l}[w(x,t,z)]&=\mathop{\hbox{Res}}\limits_{z=z_l}[m(x,t,z)]X_+^{-1}(z_l,t)\\
	&=c_l\hbox{e}^{i(\lambda_{-,l}+\lambda_{+,l})x}(0,m_1(x,t,z_l))X_+^{-1}(z_l,t),\quad l=1,2,\cdots,W,
	\end{split}
\end{equation}
where   subscript $l$ represent the $l$th column of the matrix. In particular, we can express the residue conditions for $w(x,t,z)$ in the following:
\begin{equation}
	\begin{split}
	\mathop{\hbox{Res}}\limits_{z=z_l}[w_1(x,t,z)-\frac{ip_+^*}{z+\lambda_+(z)}w_2(x,t,z)]=&0,\\
		\mathop{\hbox{Res}}\limits_{z=z_l}[w_2(x,t,z)+\frac{ip_+}{z+\lambda_+(z)}w_1(x,t,z)]=&c_l\hbox{e}^{ix(\lambda_{-,l}+\lambda_{+,l})}\\
		&\times  (w_1(x,t,z_l)-\frac{ip_+^*}{z_l+\lambda_{+,l}}w_2(x,t,z_l)).
	\end{split}
\end{equation}

Solving the  RHP  for $w$, we have
\begin{equation}
	\begin{split}
		w=I+\sum_{l=1}^W\frac{1}{z-z_l}	\mathop{\hbox{Res}}\limits_{z=z_l}[w(x,t,z)]-\frac{1}{2\pi i}\int_{\Xi_+}\frac{[w^-(I-\tilde{J})](x,t,\zeta)}{\zeta-z}d\zeta,\quad z\in\mathbb{C}\backslash\Xi_+.
		\end{split}
\end{equation}
From \eqref{gl-1}, \eqref{n-4} and \eqref{gl-2}, a direct computation shows
\begin{equation}\label{dengy-2}
	\begin{split}
		m=&X_++\sum_{l=1}^W\frac{1}{z-z_l}\mathop{\hbox{Res}}\limits_{z=z_l}[m(x,t,z)]X_+^{-1}(z_l,t)X_+\\
		&-\frac{1}{2\pi i}\int_{\Xi_+}\frac{[m^-(X_+-I)J_0X_+^{-1}](x,t,\zeta)}{\zeta-z}X_+d\zeta,\quad z\in\mathbb{C}\backslash\Xi_+.
	\end{split}
\end{equation}
Considering \eqref{dengy-1} and  \eqref{dengy-2}, we have 
\begin{equation}\label{dengy-3}
	\begin{split}
		&(1+\frac{ip_+^*}{2\lambda_{+,l}(z_l+\lambda_{+,l})}c_l\hbox{e}^{i(\lambda_{-,l}+\lambda_{+,l})x})m_1(x,t,z_l)\\
		=&(I-\frac{1}{2\pi i}\int_{\Xi_+}
		\frac{[m^-(X_+-I)J_0X_+^{-1}](x,t,\zeta)}{\zeta-z_l}d\zeta\\
		&+\sum_{l'=1,l'\neq l}^W\frac{1}{z_l-z_{l'}}c_{l'}\hbox{e}^{i(\lambda_{-,l'}+\lambda_{+,l'})x}(0,m_1(x,t,z_{l'}))X_+^{-1}(z_{l'},t))X_{+,1}(z_l,t),\\
		&l=1,2,\cdots,W.
	\end{split}
\end{equation} 
 By solving the  \eqref{dengy-2} and \eqref{dengy-3} (together with \eqref{dengy-1}), one can determine the  solution of the RHP.  We recover the potential as follows:
\begin{equation}\label{dengy-4}
	p^*=2i\lim\limits_{z\to\infty  \atop \rm{Im}z>0 }zm_{21}(x,t,z).
\end{equation}	
Now by the expression \eqref{dengy-2} of the $m$, we have
\begin{equation}\label{dengy-5}
	\begin{split}
		m=&I+\frac{i}{2z}\sigma_3P_++\frac{1}{z}\sum_{l=1}^Wc_l\hbox{e}^{i(\lambda_{-,l}+\lambda_{+,l})x}(0,m_1(x,t,z_l)) X_+^{-1}(z_l,t)\\\
	&+\frac{1}{2\pi i z}\int_{\Xi_+}{[m^-(X_+-I)J_0X_+^{-1}](x,t,\zeta)}d\zeta+O(\frac{1}{z^2}),\quad z\to\infty\wedge \hbox{Im}z>0.
	\end{split}
\end{equation}
Finally, using   \eqref{dengy-4} and \eqref{dengy-5},    the solution of the defocusing Hirota equation with asymmetric NZBCs is given by
\begin{equation}
	\begin{split}
		p^*=&p_+^*(1-\sum_{l=1}^W\frac{c_l}{\lambda_{+,l}}c_l\hbox{e}^{i(\lambda_{-,l}+\lambda_{+,l})x}m_{21}(x,t,z_l))\\
		&
	+\frac{1}{2\pi i}\int_{\Xi_-\cup\Xi_\circ}	\frac{1}{\lambda_+(\zeta)}[(\frac{ip_+^*}{\zeta+\lambda_+(\zeta)}J_{0,12}(x,t,\zeta)+J_{0,11}(x,t,\zeta))p_+^*m_{22}^-(x,t,\zeta)\\
		&-(\frac{ip_+^*}{\zeta+\lambda_+(\zeta)}J_{0,22}(x,t,\zeta)+J_{0,21}(x,t,\zeta))p_+m_{21}^-(x,t,\zeta)]d\zeta.
	\end{split}
\end{equation}

	\section*{Acknowledgements}
	
	This work is supported by the National Natural Science Foundation of China (Nos. 11371326  and 12271488).

\end{document}